\documentclass[12pt,aps,final,showpacs]{revtex4}

\def\vb#1{\mbox{\boldmath$#1$}}
\def\pd#1#2{\frac{\partial #1}{\partial #2}}

\def\wh#1{\widehat{#1}}
\def\bdot{\,\vb{\cdot}\,}
\def\btimes{\,\vb{\times}\,}

\def\bhat{\wh{{\sf b}}}
\def\bstar{{\sf b}^{*}}
\def\x{\mbox {\boldmath $\xi$}}
\def\L{\mathcal L}
\def\K{\mathcal K}
\def\P{\mathcal P}

\begin{document}

\title{Energy-conserving Finite-$\beta$ Electromagnetic Drift-fluid Equations}

\author{A.~J.~Brizard}
\affiliation{Department of Chemistry and Physics, Saint Michael's College \\ Colchester, VT 05439, USA}

\begin{flushright}
June 7, 2005
\end{flushright}

\begin{abstract}
Nonlinear energy-conserving drift-fluid equations that are suitable to describe self-consistent finite-$\beta$ low-frequency electromagnetic (drift-Alfv\'{e}n) turbulent fluctuations in a nonuniform, anisotropic, magnetized plasma are derived from a variational principle. The variational principle is based on a drift-fluid Lagrangian that contains linear and nonlinear $E\times B$ velocities derived directly from the corresponding single-particle finite-$\beta$ gyrocenter  Hamiltonian (in the zero-Larmor-radius limit). Covariant electromagnetic effects introduce a magnetic generalization to the standard ion polarization density as well as introduce a new ion magnetization current, which are both missing from existing gyrofluid and drift-fluid Poisson-Amp\`{e}re equations. An exact energy conservation law is also derived directly from the drift-fluid Lagrangian by application of the Noether method.
\end{abstract}


\maketitle

\section{\label{sec:intro}Introduction}

The derivation of dissipationless nonlinear reduced-fluid equations that explicitly conserve energy represents an important goal of the numerical simulation communities in fusion \cite{Red_Fus_1}-\cite{Red_Fus_8} and space 
\cite{redfluid_space_1}-\cite{redfluid_space_2} plasma physics. The existence of an exact energy conservation law plays an important role in benchmarking numerical investigations of energy-transfer processes in a nonlinear plasma turbulent state between the fluid kinetic energy, its internal energy, and the electromagnetic field energy. Hence, if and when dissipative effects and/or advanced fluid-closure schemes are introduced in a nonlinear dissipationless reduced-fluid model, one can develop a deeper understanding of the roles played by dissipationless and dissipative effects on the long-time evolution of the turbulent plasma state.

The variational formulation of dissipationless fluid dynamics ensures that the nonlinear fluid equations satisfy an exact energy conservation law derived by application of the Noether method \cite{Noether_1}-\cite{Noether_3}. Previous variational formulations of reduced fluid models include the drift-fluid work of Pfirsch and Correa-Restrepo \cite{PCR_1,PCR_2} and the recent gyrofluid works of Strintzi and Scott \cite{SS} and Strintzi, Scott, and Brizard 
\cite{SSB}. In contrast to Refs.~\cite{SS,SSB}, however, we ignore finite-Larmor-radius (FLR) effects to focus our attention on the drift-fluid polarization and magnetization effects introduced by the low-frequency magnetic field fluctuations.

In the energy-conserving drift-Alfv\'{e}n model presented here, the perturbed electric and magnetic fields
\begin{equation}
{\bf E} \;\equiv\; -\,\nabla\phi \;-\; \frac{\bhat_{0}}{c}\;\pd{A_{\|}}{t} \;\;\;{\rm and}\;\;\; {\bf B}_{\bot} 
\;\equiv\; \nabla A_{\|}\btimes\bhat_{0},
\label{eq:E_B}
\end{equation}
are expressed in terms of the fluctuation scalar potentials $\phi$ and $A_{\|}$, and the time-independent background magnetic field ${\bf B}_{0} = B_{0}\,\bhat_{0}$ is assumed to be spatially nonuniform. From Eq.~(\ref{eq:E_B}), we define the following perturbed $E\times B$ velocities:
\begin{eqnarray}
{\bf u}_{E} & \equiv & \frac{c\bhat_{0}}{B_{0}}\btimes\nabla_{\bot}\phi \;=\; {\bf E}_{\bot}\btimes\frac{c\bhat_{0}}{B_{0}}, \label{eq:ExB_lin} \\
V_{\|} & \equiv & \frac{c\,\nabla_{\bot}\phi}{B_{0}}\bdot\frac{\nabla_{\bot}A_{\|}}{B_{0}} \;=\; \frac{c\bhat_{0}}{B_{0}^{2}}\bdot\left({\bf E}_{\bot}\btimes{\bf B}_{\bot}\right), \label{eq:ExB_nonlin}
\end{eqnarray}
where ${\bf u}_{E}$ denotes the linear perturbed $E\times B$ velocity and $V_{\|}$ denotes the parallel component of the nonlinear perturbed $E\times B$ velocity. Lastly, according to the finite-$\beta$ ordering \cite{finite_beta_1}-\cite{finite_beta_3} ($m_{e}/m_{i} < \beta \ll 1$), we neglect the perpendicular component of the fluctuating vector potential 
(${\bf A}_{\bot} \equiv 0$), so that the parallel component of the fluctuating magnetic field is assumed to vanish in what follows.

The remainder of this paper is organized as follows. In Sec.~\ref{sec:dfld}, we present the heuristic derivation of the nonlinear drift-fluid Lagrangian density to be used, in Sec.~\ref{sec:dfde}, in the variational derivation of our self-consistent, energy-conserving drift-Alfv\'{e}n equations. In Sec.~\ref{sec:dfecl}, we use the Noether method to derive the local form of the energy conservation law for our drift-Alfv\'{e}n model. Next, we integrate the local form of the energy conservation law to obtain its global form, in which the energy-transfer processes between the parallel kinetic energy, internal energy, and electromagnetic field energy are manifestly expressed. In Sec.~\ref{sec:disc}, we summarize our variational derivation  of the energy-conserving drift-Alfv\'{e}n model. Lastly, in Appendix A, we present a procedure developed for including heat fluxes within a fluid Lagrangian variational principle, while in Appendix B, we briefly outline the diamagnetic-cancellation procedure involving the addition of energy-conserving gyroviscous terms in the drift-fluid equations for $(u_{\|},p_{\bot},p_{\|})$.

\section{\label{sec:dfld}Drift-fluid Lagrangian Density}

The four-moment drift-fluid Lagrangian for the present anisotropic-temperature finite-$\beta$ electromagnetic model is based on the previous electrostatic gyrofluid work of Strintzi, Scott, and Brizard \cite{SSB}. Here, the drift-fluid Lagrangian density (for a fluid species of mass $m$ and charge $e$) is defined as
\begin{equation}
\L_{{\rm df}} \;\equiv\; \frac{mn}{2}\;u_{\|}^{2} \;-\; \left( p_{\perp} \;+\; \frac{p_{\|}}{2} \right) \;+\; en \left(
{\bf A}_{0} \;+\; A_{\|}\,\bhat_{0}\right) \bdot \frac{{\bf u}}{c} \;-\; en\;\phi \;-\; \L_{{\rm ZLR}}\left[\; 
H_{2{\rm gy}} \;\right],
\label{eq:a_gy}
\end{equation}
where $u_{\|} \equiv {\bf u}\bdot\bhat_{0}$ denotes the drift-fluid velocity parallel to the background magnetic field and all terms (except the last one) are standard fluid Lagrangian terms. The last term represents the zero-Larmor-radius (ZLR) limit of the electromagnetic nonlinearities contained in the second-order finite-$\beta$ gyrocenter Hamiltonian \cite{HLB_88}: 
\begin{eqnarray*} 
H_{2{\rm gy}} & = & -\; \frac{mc^{2}}{2\,B_{0}^{2}}\;|\nabla_{\bot}\phi|^{2} \;+\; mv_{\|}\;
\frac{c\,\nabla_{\bot}\phi}{B_{0}}\bdot\frac{\nabla_{\bot}A_{\|}}{B_{0}} \;-\; \frac{m\,v_{\|}^{2}}{2}\;
\frac{|\nabla_{\bot}A_{\|}|^{2}}{B_{0}^{2}} \\
 & \equiv & -\;\frac{m}{2}\,\left(|{\bf u}_{E}|^{2} \;+\; v_{\|}^{2}\; \frac{|{\bf B}_{\bot}|^{2}}{B_{0}^{2}} \right) 
\;+\; mv_{\|}\,V_{\|},
\end{eqnarray*}
where $v_{\|}$ denotes the parallel gyrocenter particle velocity, the $E\times B$ velocities ${\bf u}_{E}$ and $V_{\|}$ are defined in Eqs.~(\ref{eq:ExB_lin}) and (\ref{eq:ExB_nonlin}), respectively, and the background magnetic-field nonuniformity was neglected. By heuristically transforming this second-order gyrocenter Hamiltonian to its drift-fluid version, following a procedure outlined in Refs.~\cite{PCR_1,PCR_2}, the drift-fluid Lagrangian density (\ref{eq:a1}) for each fluid species becomes
\begin{eqnarray}
\L_{{\rm df}} & = & \frac{mn}{2}\;\left|u_{\|}\,(\bhat_{0} + {\bf B}_{\bot}/B_{0}) + {\bf u}_{E}\right|^{2} \;-\; 
\left( p_{\perp} \;+\; \frac{p_{\|}}{2} \right) \;+\; en \left({\bf A}_{0} \;+\; A_{\|}\,\bhat_{0}\right) \bdot 
\frac{{\bf u}}{c} \;-\; e\,n\;\phi \nonumber \\
 & \equiv & \frac{mn}{2}\,U_{\|}^{2} \;+\; en\,{\bf A}_{0}\bdot\frac{{\bf u}}{c} \;-\; 
\left( p_{\bot} + \frac{p_{\|}}{2} \right) \;-\; en \left( \Phi^{*} \;-\; \frac{u_{\|}}{c}\,A_{\|}^{*} \right),
\label{eq:a1}
\end{eqnarray}
where we introduced the definitions
\begin{equation}
\left. \begin{array}{rcl}
U_{\|} & \equiv & u_{\|}\,b \;\equiv\; u_{\|}\;\left(1 \;+\; |{\bf B}_{\bot}|^{2}/B_{0}^{2}\right)^{1/2} \\
 &  & \\
e\,\Phi^{*} & \equiv & e\,\phi \;-\; m|{\bf u}_{E}|^{2}/2 \\
 &  & \\
(e/c)\,A_{\|}^{*} & \equiv & (e/c)\,A_{\|} \;-\; m\,V_{\|}
\end{array} \right\},
\label{eq:phi_Ab}
\end{equation}
with the effective potentials $\Phi^{*}$ and $A_{\|}^{*}$ both containing linear and nonlinear field terms. We point out that, while $u_{\|}$ denotes the parallel drift-fluid velocity along the unperturbed (background) magnetic field 
${\bf B}_{0}$, $U_{\|}$ denotes the parallel drift-fluid velocity along the total magnetic field ${\bf B}_{0} + 
{\bf B}_{\bot}$. Moreover, although the term $b^{2} - 1$ is considered small, it is kept in the drift-fluid Lagrangian 
(\ref{eq:a1}) to allow a covariant treatment of drift-fluid polarization and magnetization effects in our model [see 
Eqs.~(\ref{eq:E_pol})-(\ref{eq:B_mag}) below].

We note that the fluid velocity ${\bf u}$ appearing in the drift-fluid Lagrangian density (\ref{eq:a1}) is defined as the particle-fluid velocity expressed in terms of moments of the gyroangle-independent gyrocenter distribution function 
$F({\bf R},v_{\|},\mu,t)$ defined in Ref.~\cite{Brizard_89}:
\begin{eqnarray} 
n({\bf r},t)\,{\bf u}({\bf r},t) & = & \int\,f({\bf x},{\bf v},t)\,{\bf v}\,\delta^{3}({\bf x} - {\bf r})\;d^{3}x\,d^{3}v 
\label{eq:particle} \\
 & \equiv & \int\,F({\bf R},v_{\|},\mu,t)\; \left\langle ({\bf v}_{\bot} + {\bf v}_{{\rm gc}} + \cdots)\,
\delta^{3}({\bf R} + \vb{\rho} - {\bf r})\right\rangle \;d^{6}Z. \label{eq:gyro}
\end{eqnarray}
Here, $\langle\cdots\rangle$ denotes a gyroangle average and we used the transformation from a particle-fluid moment 
(\ref{eq:particle}) to a gyrofluid moment (\ref{eq:gyro}), which was introduced in Ref.~\cite{brizard}, where 
${\bf v}_{{\rm gc}}$ denotes the guiding-center (gc) velocity and higher-order corrections (e.g., electromagnetic fluctuations) are omitted for simplicity. Furthermore, because of the presence of $\delta^{3}({\bf R} + \vb{\rho} - 
{\bf r})$ in Eq.~(\ref{eq:gyro}), the gyrofluid moment of the perpendicular particle velocity ${\bf v}_{\bot}$ does not vanish but, instead, yields the divergenceless term
\[ \int\,F({\bf R},v_{\|},\mu,t)\; \left\langle {\bf v}_{\bot}\,\delta^{3}({\bf R} + \vb{\rho} - {\bf r})\right\rangle 
\;d^{6}Z \;=\; -\;\nabla\btimes\left( p_{\bot}\;\frac{c\bhat_{0}}{eB_{0}} \right) \]
which introduces an important diamagnetic contribution to the particle flux (\ref{eq:gyro}):
\begin{equation} 
n\,{\bf u}_{{\rm P}} \;\equiv\; \frac{c\bhat_{0}}{eB_{0}}\btimes\left( p_{\bot}\,\nabla\ln B_{0} \;+\; p_{\|}\,\bhat_{0}\bdot\nabla\bhat_{0} \right) \;-\; \nabla\btimes\left( p_{\bot}\;\frac{c\bhat_{0}}{eB_{0}} \right) \;=\; 
\frac{c\bhat_{0}}{eB_{0}}\btimes \left( \nabla\bdot{\sf P} \right),
\label{eq:u_CGL}
\end{equation}
where ${\sf P}$ denotes the Chew-Goldberger-Low (CGL) pressure tensor \cite{CGL} [see Eq.~(\ref{eq:CGL}) below]. 

The appearance of the perturbed electric and magnetic fields in the drift-fluid Lagrangian (\ref{eq:a1}) yields the following expressions for the drift-fluid polarization and magnetization vectors:
\begin{eqnarray}
{\bf P}_{\bot} & \equiv & \pd{\L_{{\rm df}}}{{\bf E}_{\bot}} \;=\; \sum\;\frac{mnc^{2}}{B_{0}^{2}} \left( {\bf E}_{\bot} \;+\; \frac{u_{\|}}{c}\;
\bhat_{0}\btimes{\bf B}_{\bot} \right), \label{eq:E_pol} \\
{\bf M}_{\bot} & \equiv & \pd{\L_{{\rm df}}}{{\bf B}_{\bot}} \;=\; \sum\; mn\,\frac{u_{\|}}{B_{0}} \left( u_{\|}\;\frac{{\bf B}_{\bot}}{B_{0}} \;+\; 
{\bf E}_{\bot}\btimes\frac{c\bhat_{0}}{B_{0}} \right). \label{eq:B_mag}
\end{eqnarray}
Because of the mass dependence in the drift-fluid polarization and magnetization vectors (\ref{eq:E_pol})-(\ref{eq:B_mag}), the drift-fluid polarization and magnetization effects are especially important for ion fluid species (in a quasineutral plasma).

The total Lagrangian density for our four-moment electromagnetic drift-fluid model is, therefore, defined as the sum of the Lagrangian density of the electromagnetic field  and the drift-fluid Lagrangian density (\ref{eq:a1}) for each fluid species (summation over species is implied):
\begin{eqnarray}
\L & = & \frac{1}{8\pi} \left( \left|{\bf E}_{\bot}\right|^{2} \;-\; |{\bf B}_{\bot}|^{2} \right) \;+\; 
en \left( {\bf A}_{0}\bdot\frac{{\bf u}}{c} \;+\; \frac{u_{\|}}{c}\,A_{\|}^{*} \;-\; \Phi^{*} \right) \nonumber \\
 &  &\mbox{}+\; \frac{mn}{2}\;U_{\|}^{2} \;-\; \left( p_{\perp} \;+\; \frac{p_{\|}}{2} \right),
\label{eq:Lag_total}
\end{eqnarray}
where we have omitted the contribution from the parallel electric field in the electric field energy (i.e., 
$|{\bf E}|^{2} \rightarrow |{\bf E}_{\bot}|^{2}$), which removes the parallel displacement current $(\partial_{t}E_{\|})$ in the parallel drift-fluid Amp\`{e}re equation [see Eq.~(\ref{eq:A_Maxwell})]. The variational fields are the four drift-fluid moments $(n,{\bf u},p_{\bot},p_{\|})$ for each fluid species and the electromagnetic potentials $(\phi,A_{\|})$. Here, we note that only the first four drift-fluid moments $(n,{\bf u},p_{\bot},p_{\|})$ appear in the drift-fluid Lagrangian density (\ref{eq:a1}) and higher-order drift-fluid moments (e.g., heat fluxes) are omitted as variational dynamical fields; future work will consider the introduction of higher-order drift-fluid moments in a Lagrangian variational setting \cite{thermal_momentum} (see Appendix A for further comments). Lastly, we note that, because the background magnetic field ${\bf B}_{0}$ is independent of the variational fields, the magnetic-energy term $-\,|{\bf B}_{0}|^{2}/8\pi$ has been removed from the Lagrangian density (\ref{eq:Lag_total}).

\section{\label{sec:dfde}Variational Derivation of Drift-fluid Dynamical Equations}

The drift-Alfv\'{e}n variational principle
\begin{equation}
\int\; \delta\L(n,{\bf u},p_{\|},p_{\bot}; \phi, A_{\|}; {\bf x})\;\; d^{3}x\,dt \;=\; 0
\label{eq:VPDF}
\end{equation}
associated with the drift-fluid Lagrangian density (\ref{eq:Lag_total}) must be expressed in terms of the Eulerian variations $(\delta n,\delta{\bf u},\delta p_{\bot},\delta p_{\|}; \delta\phi, \delta A_{\|})$: 
\begin{eqnarray}
\delta\L & \equiv & \left( \delta n\;\pd{\L}{n} + \delta{\bf u}\bdot\pd{\L}{{\bf u}} + \delta p_{\|}\;\pd{\L}{p_{\|}} +
\delta p_{\bot}\;\pd{\L}{p_{\bot}} + \delta\phi\;\pd{\L}{\phi} + \delta A_{\|}\;\pd{\L}{A_{\|}} \right) \nonumber \\
 &  &\mbox{}+\; \nabla\delta\phi\bdot\pd{\L}{(\nabla\phi)} \;+\; \nabla\delta A_{\|}\bdot\pd{\L}{(\nabla A_{\|})},
\label{eq:Lag_var}
\end{eqnarray}
where the additional ${\bf x}$-dependence in $\L(\cdots;{\bf x})$ arises from the nonuniform background magnetic field 
${\bf B}_{0}$. Here, the Eulerian variations for the drift-fluid moments $(\delta n,\delta{\bf u},\delta p_{\bot},\delta 
p_{\|})$ are not independent of each other but are instead expressed in terms of a virtual fluid displacement $\x$ (for each fluid species). 

The derivation of the Eulerian variations $(\delta n,\delta p_{\bot},\delta p_{\|})$ are based on dynamical constraints for the drift-fluid moments $\eta^{a} = (n,p_{\bot},p_{\|})$. The first dynamical constraint is associated with mass conservation, expressed in terms of the continuity equation
\begin{equation}
\frac{dn}{dt} \;+\; n\;\nabla\bdot{\bf u} \;=\; 0, 
\label{eq:continuity}
\end{equation}
where $d/dt = \partial/\partial t + {\bf u}\bdot\nabla$ denotes the total time derivative. The next two dynamical constraints are associated with the conservation of the first two single-particle adiabatic invariants, expressed in terms of the Chew-Goldberger-Low equations for the perpendicular and parallel pressures \cite{CGL}:
\begin{eqnarray}
\frac{dp_{\bot}}{dt} \;+\; p_{\bot}\;\nabla\bdot{\bf u} \;+\; p_{\perp}\;({\bf I} - \bhat_{0}\bhat_{0}):\nabla{\bf u} 
& = & 0, \label{eq:pperp_CGL} \\
\frac{dp_{\|}}{dt} \;+\; p_{\|}\;\nabla\bdot{\bf u} \;+\; 2\,p_{\parallel}\;\bhat_{0}\bhat_{0}:\nabla{\bf u} & = & 0, \label{eq:ppar_CGL}
\end{eqnarray}
where the higher-order heat-flux moments ${\bf q}_{\bot}^{(\|)}$ and ${\bf q}_{\bot}^{(\bot)}$ are omitted here but are considered later in Appendix B.

We begin our derivation of expressions for the Eulerian variations to be used in Eq.~(\ref{eq:Lag_var}) by transforming the dynamical constraints (\ref{eq:continuity}) and (\ref{eq:pperp_CGL})-(\ref{eq:ppar_CGL}) into Lagrangian variations $\Delta\eta^{a} = (\Delta n, \Delta p_{\bot}, \Delta p_{\|})$ by introducing the following limiting process
\begin{equation}
\lim_{\Delta t \rightarrow 0} \left( \Delta t\;\frac{d\eta^{a}}{dt} \right) \;\equiv\; \Delta\eta^{a} 
\;\;\;{\rm and}\;\;\; \lim_{\Delta t \rightarrow 0}\;\left( {\bf u}\,\Delta t\right) \;\equiv\; \x
\label{eq:Lagvar_def}
\end{equation}
where $\x$ denotes the virtual fluid-displacement (for each fluid species), with the Lagrangian variation of the fluid velocity is defined as 
$\Delta{\bf u} \equiv d\x/dt$. Hence, using the procedure (\ref{eq:Lagvar_def}) on the constraint equations (\ref{eq:continuity})-(\ref{eq:ppar_CGL}), we find the Lagrangian variations 
\begin{eqnarray*} 
\Delta n & = & -\,n\;\nabla\bdot\x, \\ 
\Delta p_{\bot} & = & -\,p_{\bot}\;\nabla\bdot\x - p_{\bot}\;({\bf I} - \bhat_{0}\bhat_{0}):\nabla\x, \\ 
\Delta p_{\|} & = & -\,p_{\|}\;\nabla\bdot\x - 2\,p_{\bot}\;\bhat_{0}\bhat_{0}:\nabla\x.
\end{eqnarray*}
Next, using the relation $\delta\eta^{a} \equiv \Delta\eta^{a} - \x\bdot\nabla\eta^{a}$ between Lagrangian and Eulerian variations \cite{Newcomb}, the Eulerian variations $(\delta n, \delta{\bf u}, \delta p_{\|}, \delta p_{\bot})$ to be used in the variational principle (\ref{eq:Lag_var}) are now defined as 
\begin{equation}
\left. \begin{array}{rcl}
\delta n & = & -\;\nabla \bdot (n\;\x) \\
 &  & \\
\delta {\bf u} & = & \partial_{t}\x \;+\; ({\bf u}\bdot \nabla )\x \;-\; (\x \bdot \nabla){\bf u} \\
 &  & \\
\delta p_{\|} & = & -\,\nabla\bdot( p_{\|}\,\x) \;-\; 2p_{\|}\; \bhat_{0}\bhat_{0}:\nabla\x \\
 &  & \\
\delta p_{\bot} & = & -\,\nabla \bdot(p_{\bot}\;\x ) \;-\; p_{\bot}\; ({\bf I} - \bhat_{0}\bhat_{0}):\nabla\x
\end{array} \right\}.
\label{eq:constraints}
\end{equation}

The remaining drift-fluid equations include the evolution equation for the parallel drift-fluid velocity $u_{\|}$ and an explicit expression for the perpendicular component of the drift-fluid velocity ${\bf u}$. Lastly, a self-consistent treatment involves the appropriate Poisson and parallel Amp\`{e}re equations for the potentials $(\phi,A_{\|})$ expressed in terms of the drift-fluid moments $(n,u_{\|})$.

\subsection{Variation of the Lagrangian density}

We now insert the Eulerian variations (\ref{eq:constraints}) into the drift-fluid Lagrangian variation 
(\ref{eq:Lag_var}), and rearranging several terms to isolate $\x$ and $\delta\psi^{i} = (\delta\phi,\delta A_{\|})$, to obtain
\begin{eqnarray}
\delta\L & \equiv & -\; \x \bdot\left[\; \pd{}{t}\left(\pd{\L}{{\bf u}}\right) \;+\; \nabla\bdot\left({\bf u}\;
\pd{\L}{{\bf u}} \right) \;+\; \nabla{\bf u}\bdot\pd{\L}{{\bf u}} \;+\; \nabla\bdot{\sf P} \;-\; 
\left( \eta^{a}\;\nabla\pd{\L}{\eta^{a}} \right) \;\right] \nonumber \\
 &  &\mbox{}+\; \delta\psi^{i} \left[\; \pd{\L}{\psi^{i}} - \nabla\bdot\pd{\L}{(\nabla\psi^{i})} \;\right] \;+\; 
\pd{\Lambda}{t} \;+\; \nabla\bdot\vb{\Gamma},
\label{eq:Eul_Lag}
\end{eqnarray}
where 
\[ \eta^{a}\;\nabla\pd{\L}{\eta^{a}} \;\equiv\; n\;\nabla\pd{\L}{n} + p_{\|}\;\nabla\pd{\L}{p_{\|}} + p_{\bot}\;
\nabla\pd{\L}{p_{\bot}}, \]
and the tensor ${\sf P}$ denotes the CGL pressure tensor \cite{CGL}
\begin{equation}
{\sf P} \;\equiv\; -\;p_{\bot}\,\pd{\L}{p_{\bot}}\;({\bf I} - \bhat_{0}\bhat_{0}) \;-\; 2p_{\|}\,\pd{\L}{p_{\|}}\;
\bhat_{0}\bhat_{0} \;=\; p_{\bot}\;({\bf I} - \bhat_{0}\bhat_{0}) \;+\; p_{\|}\;\bhat_{0}\bhat_{0}.
\label{eq:CGL}
\end{equation}
While the space-time divergence terms $\partial_{t}\Lambda + \nabla\bdot\vb{\Gamma}$ in Eq.~(\ref{eq:Eul_Lag}), expressed in terms of the Noether fields
\begin{equation}
\Lambda \;\equiv\; \x\bdot\pd{\L}{{\bf u}}, \label{eq:Lambda}
\end{equation}
and
\begin{equation}
\vb{\Gamma} \;\equiv\; {\bf u}\;\left(\x\bdot\pd{\L}{{\bf u}}\right) \;+\; \delta\psi^{i}\;\pd{\L}{(\nabla\psi^{i})} 
\;+\; \left[\; {\sf P} \;-\; \left( \eta^{a}\;\pd{\L}{\eta^{a}} \right) {\bf I} \;\right]\bdot\x. 
\label{eq:Gamma}
\end{equation}
do not play a role in the variational principle $\int \delta\L\,d^{3}x\,dt = 0$, they play a crucial role in the derivation of exact conservation laws based on the Noether method \cite{Noether_1}-\cite{Noether_3} (see Sec.~\ref{sec:dfecl}).

\subsection{Drift-fluid velocity equations}

The stationarity of the action functional $\int\L\,d^{3}x\,dt$ with respect to a arbitrary virtual fluid displacement 
$\x$ yields the Euler-Poincar\'{e} equation  \cite{EP_eq} (for each fluid species)
\begin{equation}
0 \;=\; \pd{}{t}\left(\pd{\L}{{\bf u}}\right) \;+\; \nabla\bdot\left({\bf u}\;
\pd{\L}{{\bf u}} \right) \;+\; \nabla{\bf u}\bdot\pd{\L}{{\bf u}} \;+\; \nabla\bdot{\sf P} \;-\; 
\left( \eta^{a}\;\nabla\pd{\L}{\eta^{a}} \right),
\label{eq:EP}
\end{equation}
which describes the time evolution of the drift-fluid velocity ${\bf u}$.  Upon substituting derivatives of the drift-fluid Lagrangian density 
(\ref{eq:Lag_total}) into the Euler-Poincar\'{e} equation (\ref{eq:EP}), and using the fact that the background magnetic field ${\bf B}_{0}$ is assumed to be a time-independent nonuniform vector field, Eq.~(\ref{eq:EP}) can be written as
\begin{equation}
0 \;=\; n\,\bhat_{0}\;\pd{}{t} \left( mu_{\|}\,b^{2} \;+\; \frac{e}{c}\,A_{\|}^{*} \right) \;-\; \frac{en}{c}\;{\bf u}\btimes{\bf B}^{*} \;+\; 
\nabla\bdot{\sf P} \;+\; n\;\nabla\left[\; e\,\Phi^{*} \;+\; \frac{m}{2}\,(u_{\|}b)^{2} \;\right],
\label{eq:gyro_velocity}
\end{equation}
where we have introduced the following divergenceless field
\begin{eqnarray}
{\bf B}^{*} & \equiv & {\bf B}_{0} \;+\; \nabla \btimes\left[\; \left( A_{\|}^{*} \;+\; \frac{mc}{e}\,u_{\|}\,b^{2} 
\right) \bhat_{0} \;\right], \label{eq:Bstar_def}
\end{eqnarray}
from which we define
\begin{eqnarray}
B_{\|}^{*} & \equiv & \bhat_{0}\bdot{\bf B}^{*} \;=\; B_{0} \;+\; \left[\; A_{\|}^{*} \;+\; \left(\frac{mc}{e}\right)\,
u_{\|}\,b^{2} \;\right] \bhat_{0}\bdot\nabla\btimes\bhat_{0}, \label{eq:Bpar_star} \\
\bstar & \equiv & \frac{{\bf B}^{*}}{B_{\|}^{*}} \;=\; \bhat_{0} \;+\; \frac{c\bhat_{0}}{eB_{\|}^{*}}\btimes\left[\; 
\left( mu_{\|}\,b^{2} \;+\; \frac{e}{c}\,A_{\|}^{*} \right) \vb{\kappa}_{0} \;-\; \nabla\left( mu_{\|}\,b^{2} \;+\; 
\frac{e}{c}\,A_{\|}^{*} \right) \;\right], \label{eq:bBstar_def}
\end{eqnarray}
where $\vb{\kappa}_{0} = \bhat_{0}\bdot\nabla\bhat_{0}$ denotes the background magnetic curvature. We note that the term 
$\bhat_{0}\bdot\nabla\btimes\bhat_{0}$ in Eq.~(\ref{eq:Bpar_star}), which is related to the unperturbed parallel current flowing along the background magnetic field, may vanish for some magnetic geometries (e.g., magnetic-dipole geometry).

The Euler-Poincar\'{e} equation (\ref{eq:gyro_velocity}) can be divided into two equations: one equation that expresses the drift-fluid velocity ${\bf u}$ in terms of the drift-fluid moments $(n,u_{\|},p_{\bot},p_{\|})$ and the scalar potentials $(\phi,A_{\|})$, and one equation that describes the time evolution of the parallel drift-fluid velocity 
$u_{\|}$. 

\subsubsection{Drift-fluid velocity}

The first equation is obtained by taking the cross-product of Eq.~(\ref{eq:gyro_velocity}) with $\bhat_{0}$, which yields the following first-order expression for the drift-fluid velocity:
\begin{equation}
{\bf u} \;\equiv\; u_{\|}\,\bstar \;+\; \frac{c\bhat_{0}}{en\,B_{\|}^{*}}\btimes \left[\; \nabla\bdot{\sf P} \;+\; n\;\nabla\left(e\,\Phi^{*} \;+\; \frac{m}{2}\,(u_{\|}b)^{2}\,\right) \;\right],
\label{eq:u_vector}
\end{equation}
where $\bstar$ is defined in Eq.~(\ref{eq:bBstar_def}). Here, the divergence of the CGL pressure tensor 
(\ref{eq:CGL}) is
\begin{equation}
\nabla\bdot{\sf P} \;=\; \nabla p_{\bot} \;+\; p_{\Delta}\;\vb{\kappa}_{0} \;+\; \left[\nabla\bdot\left(p_{\Delta}\,
\bhat_{0}\right)\right]\;\bhat_{0},
\label{eq:CGL_div}
\end{equation}
(with $p_{\Delta} \equiv p_{\|} - p_{\bot}$ denotes the pressure anisotropy) so that $\bhat_{0}\btimes(\nabla\bdot{\sf P}) 
= \bhat_{0}\btimes(\nabla p_{\bot} + p_{\Delta}\,\vb{\kappa}_{0})$. According to Eq.~(\ref{eq:u_vector}), the drift-fluid velocity ${\bf u}$, therefore, consists of the parallel velocity $u_{\|} \equiv {\bf u}\bdot\bhat_{0}$ (note that 
$\bhat_{0}\bdot\bstar \equiv 1$) and the following perpendicular drift-fluid velocities
\begin{equation}
\left. \begin{array}{rcl}
{\bf u}_{D} & \equiv & \left(c\,\bhat_{0}/en\,B_{\|}^{*}\right)\btimes\nabla p_{\bot} \\
 &  & \\
{\bf u}_{\Phi} & \equiv & \left(c\,\bhat_{0}/eB_{\|}^{*}\right)\btimes\nabla\left(e\Phi^{*} + m\,U_{\|}^{2}/2\right) \\
 &  & \\
{\bf u}_{C} & \equiv & \left(c\,\bhat_{0}/en\,B_{\|}^{*}\right)\btimes p_{\Delta}\;\vb{\kappa}_{0} 
\end{array} \right\},
\label{eq:gyro_perp}
\end{equation}
corresponding to the diamagnetic velocity, the generalized $E\times B$ velocity (which includes the gradients of $|{\bf u}_{E}|^{2}$ and $U_{\|}^{2} \equiv u_{\|}^{2}\,b^{2}$), and the curvature-drift velocity, respectively. We note that, although the higher-order polarization drift velocity 
\begin{equation} 
{\bf u}_{{\rm pol}} \;\equiv\; \frac{\bhat_{0}}{\Omega_{0}}\btimes\frac{d{\bf u}_{E}}{dt} + \cdots
\label{eq:u_pol}
\end{equation}
is not included in the drift-fluid velocity (\ref{eq:u_vector}), it is self-consistently introduced in the net current density ${\bf J}$ (through the polarization current density $\partial_{t}{\bf P}_{\bot}$) in the form of the quasineutrality condition $\nabla\bdot{\bf J} = 0$ (see Sec.~\ref{subsec:quasineutral} for more details).

\subsubsection{Evolution equation for $u_{\|}$}

The second equation to be extracted from the Euler-Poincar\'{e} equation (\ref{eq:gyro_velocity}), which describes the time evolution equation for parallel drift-fluid velocity $u_{\|}$, can be obtained by taking the dot-product of 
Eq.~(\ref{eq:gyro_velocity}) with $\bstar$:
\begin{equation}
n\;\pd{}{t} \left( mu_{\|}\,b^{2} \;+\; \frac{e}{c}\,A_{\|}^{*} \right) \;=\; -\,\bstar\bdot\left[\; \nabla\bdot{\sf P} \;+\; n\;\nabla\left(e\,\Phi^{*} \;+\; \frac{m}{2}\,U_{\|}^{2}\,\right) \;\right].
\label{eq:u_parallel}
\end{equation}
Here, the terms on the left side are 
\[ \pd{}{t} \left( u_{\|}\,b^{2}\right) \;=\; b\;\pd{U_{\|}}{t} \;+\; u_{\|}\;\left( \frac{\nabla_{\bot}A_{\|}}{B_{0}^{2}}
\bdot\nabla_{\bot}\pd{A_{\|}}{t}\right) \]
and
\[ \frac{e}{c}\;\pd{A_{\|}^{*}}{t} \;\equiv\; \frac{e}{c}\;\pd{A_{\|}}{t} \;-\; m\;\pd{V_{\|}}{t}, \]
which contains both the parallel inductive electric field and the partial time derivative of the parallel component of the nonlinear $E \times B$ velocity (\ref{eq:ExB_nonlin}); note that the parallel component of the partial time derivative of the linear $E \times B$ velocity vanishes $(\bhat_{0}\bdot\partial_{t}{\bf u}_{E} \equiv 0)$. Note also that, although the convective derivative ${\bf u}\bdot\nabla(mu_{\|}\,b^{2})$ along the drift-fluid velocity ${\bf u}$ appears to be absent in Eq.~(\ref{eq:u_parallel}), it is actually hidden in the last term $-\,mn\,U_{\|}\;\bstar\bdot\nabla U_{\|}$; this term may appear explictly in Eq.~(\ref{eq:gyro_velocity}) by redefining the divergenceless vector (\ref{eq:Bstar_def}) into a non-divergenceless vector as was done in Ref.~\cite{SSB}.

The set of drift-fluid equations of motion for the four drift-fluid moments $(n,u_{\|},p_{\bot},p_{\|})$ are, thus, given by Eqs.~(\ref{eq:continuity}),(\ref{eq:u_parallel}), and (\ref{eq:pperp_CGL})-(\ref{eq:ppar_CGL}), respectively, while the drift-fluid velocity ${\bf u}$ is defined by Eq.~(\ref{eq:u_vector}). Note that each of the drift-fluid equations 
for $(u_{\|},p_{\bot},p_{\|})$ involves the diamagnetic advection operator ${\bf u}_{D}\bdot\nabla$, which must be cancelled by the addition of so-called gyroviscous cancellations (see Appendix B for further details). Furthermore, the drift-fluid moment equations now need to be complemented by suitable evolution equations for the electromagnetic scalar potentials $(\phi,A_{\|})$ through the low-frequency version of Maxwell's equations, in which the net charge and current densities are expressed in terms of drift-fluid charge and current densities as well as polarization and magnetization terms.

\subsection{Poisson and parallel Amp\`{e}re equations}

We now derive the drift-fluid Poisson and parallel Amp\`{e}re equations from the variational principle (\ref{eq:VPDF}) as Euler-Lagrange equations for $\psi^{i} = (\phi,A_{\|})$: 
\begin{equation}
0 \;=\; \pd{\L}{\psi^{i}} \;-\; \nabla\bdot\pd{\L}{(\nabla\psi^{i})}.
\label{eq:EL_psi}
\end{equation}
where the drift-fluid polarization and magnetization effects, represented by the vectors 
(\ref{eq:E_pol})-(\ref{eq:B_mag}), are introduced naturally in these respective equations.

The drift-fluid Poisson equation for $\phi$ can be found from the Euler-Lagrange equation (\ref{eq:EL_psi}) for 
$\psi^{i} = \phi$, which can be expressed as
\begin{eqnarray}
\frac{\nabla\bdot{\bf E}_{\bot}}{4\pi} & = & \sum\; en \;+\; \nabla_{\bot}\bdot\left[\; \sum\; \frac{mnc^{2}}{B_{0}^{2}} \left( \nabla_{\bot}\phi \;-\; \frac{u_{\|}}{c}\;\nabla_{\bot}A_{\|} \right) \;\right] \nonumber \\
 & \equiv & \sum\; en \;-\; \nabla\bdot{\bf P}_{\bot},
\label{eq:polarization}
\end{eqnarray}
where summation over fluid species is shown explicitly. This equation shows the effects of the finite-$\beta$ electromagnetic generalization of the polarization density $\rho_{{\rm pol}} = -\,\nabla\bdot{\bf P}_{\bot}$, where the drift-fluid polarization vector is defined in Eq.~(\ref{eq:E_pol}).

Next, the parallel drift-fluid Amp\`{e}re equation can be found from the Euler-Lagrange equation (\ref{eq:EL_psi}) for 
$\psi^{i} = A_{\|}$, which can be expressed as
\begin{eqnarray}
-\;\frac{\nabla_{\bot}^{2}A_{\|}}{4\pi} & = & \sum\; \frac{en}{c}\;u_{\|} \;+\; \nabla\bdot\left[\; \sum\; 
\frac{mnu_{\|}c}{B_{0}^{2}} \left( \nabla_{\bot}\phi \;-\; \frac{u_{\|}}{c}\;\nabla_{\bot}A_{\|} \right) \;\right]
\nonumber \\ 
 & \equiv & \sum\; \frac{en}{c}\;u_{\|} \;+\; \nabla\bdot\left({\bf M}_{\bot}\btimes\bhat_{0}\right),
\label{eq:A_Maxwell}
\end{eqnarray}
where the absence of the parallel displacement current $\partial_{t}E_{\|}$ results from the omission of $E_{\|}^{2}$ in the drift-fluid Lagrangian density (\ref{eq:Lag_total}). We also see that the second term in the divergence on the right side of the drift-fluid Amp\`{e}re equation represents the effects of the parallel component of the magnetization current 
$\bhat_{0}\bdot{\bf J}_{{\rm mag}} \equiv c\,\bhat_{0}\bdot\nabla\btimes{\bf M}_{\bot}$, where the magnetization vector 
${\bf M}_{\bot}$ is defined in Eq.~(\ref{eq:B_mag}) and background magnetic nonuniformity is ignored. Since the polarization vector (\ref{eq:E_pol}) is perpendicular to the background magnetic field, however, the drift-fluid polarization current ${\bf J}_{{\rm pol}} \equiv \partial{\bf P}_{\bot}/\partial t$ has no parallel component 
($\bhat_{0}\bdot{\bf J}_{{\rm pol}} \equiv 0$) and, thus, it does not appear in the parallel drift-fluid Amp\`{e}re equation (\ref{eq:A_Maxwell}). 

Lastly, we note that the covariant electromagnetic polarization and magnetization effects in the drift-fluid 
Poisson-Amp\`{e}re equations are due to (a) the $\phi-A_{\|}$ symmetry of the nonlinear Lagrangian term $mnu_{\|}\,V_{\|}$ in the drift-fluid Lagrangian (\ref{eq:a1}), and (b) the nonlinear kinetic energy terms $mn (|{\bf u}_{E}|^{2} + 
U_{\|}^{2})/2$. While such low-frequency polarization and magnetization effects appear naturally in nonlinear 
finite-$\beta$ gyrokinetic theory \cite{HLB_88}, they have not appeared explicitly in previous derivations of gyrofluid equations (e.g., Ref.~\cite{brizard}).

\subsection{\label{subsec:quasineutral}Quasineutral drift-fluid dynamics}

A fundamental aspect of drift-fluid dynamics in a strongly magnetized plasma involves the fact that the time evolution of the turbulent plasma state maintains a condition of quasineutrality \cite{BDS_2003}, where $\rho = \nabla\bdot
{\bf E}_{\bot}/4\pi \equiv 0$ and, thus, according to the drift-fluid Poisson equation (\ref{eq:polarization}), the net drift-fluid charge density \cite{PCR_1}
\begin{equation}
\sum\;en \;\equiv\; \nabla\bdot{\bf P}_{\bot} 
\label{eq:rho_def}
\end{equation}
is expressed in terms of the drift-fluid polarization vector (\ref{eq:E_pol}). This drift-fluid quasineutrality condition, therefore, can be used with Eq.~(\ref{eq:polarization}) to obtain the drift-fluid charge conservation law:
\begin{equation}
0 \;=\; \nabla\bdot\left( \sum\,en\,{\bf u} \;+\; \pd{{\bf P}_{\bot}}{t} \right) \;\equiv\; \nabla\bdot{\bf J},
\label{eq:df_charge}
\end{equation}
where the net current density
\begin{equation} 
{\bf J} \;\equiv\; \sum\;en\,{\bf u} \;+\; \pd{{\bf P}_{\bot}}{t} \;+\; \nabla\btimes\vb{\chi}
\label{eq:J_def}
\end{equation}
is defined as the sum of the net drift-fluid current density ($\sum\;en\,{\bf u}$), the drift-fluid polarization current 
($\partial_{t}{\bf P}_{\bot}$), and a divergenceless term ($\nabla\btimes\vb{\chi}$). 

The vector field $\vb{\chi}$ in Eq.~(\ref{eq:J_def}) is determined by considering the parallel drift-fluid Amp\`{e}re equation (\ref{eq:A_Maxwell}) and introducing the definition for the net parallel current density:
\begin{equation}
J_{\|} \;\equiv\; -\;\frac{c}{4\pi}\;\nabla_{\bot}^{2}A_{\|} \;=\; \sum\;en\,u_{\|} \;+\; c\,\bhat_{0}\bdot\nabla\btimes
{\bf M}_{\bot},
\label{eq:Jpar_def}
\end{equation}
where the drift-fluid magnetization vector ${\bf M}_{\bot}$ is defined in Eq.~(\ref{eq:B_mag}). Hence, we find 
$\vb{\chi} \equiv c\,{\bf M}_{\bot}$ and the net current density (\ref{eq:J_def}) is now expressed as \cite{PCR_1}
\begin{eqnarray}
{\bf J} & = & \sum\;en\,{\bf u} \;+\; \pd{{\bf P}_{\bot}}{t} \;+\; c\,\nabla\btimes{\bf M}_{\bot} \nonumber \\
 & = & J_{\|}\,\bhat_{0} \;+\; \sum\;en\,{\bf u}_{\bot} \;+\; \pd{{\bf P}_{\bot}}{t} \;+\; c\,
(\nabla\btimes{\bf M}_{\bot})_{\bot},
\label{eq:Jnet_def}
\end{eqnarray}
where we have introduced the net parallel current density (\ref{eq:Jpar_def}) and $(\nabla\btimes{\bf M}_{\bot})_{\bot}$ denotes perpendicular components of the drift-fluid magnetization current.

Lastly, we note that, while the standard polarization drift velocity (\ref{eq:u_pol}) is absent from the drift-fluid velocity (\ref{eq:u_vector}), the associated drift-fluid polarization current density $\partial_{t}{\bf P}_{\bot}$ is introduced naturally into our formalism through the drift-fluid quasineutrality condition (\ref{eq:rho_def}) and the drift-fluid charge conservation law (\ref{eq:df_charge}). Furthermore, we note that the drift-fluid polarization vector 
(\ref{eq:E_pol}) can be expressed as ${\bf P}_{\bot} \equiv \sum\,en\,{\bf w}_{\bot}$ so that the drift-fluid polarization current density
\[ \pd{{\bf P}_{\bot}}{t} \;\equiv\; \sum \left[\; en\;\frac{d{\bf w}_{\bot}}{dt} \;-\; \nabla\bdot\left(en{\bf u}\,
{\bf w}_{\bot}\right) \;\right] \]
can also be expressed in terms of the total time derivative $d/dt \equiv \partial/\partial t + {\bf u}\bdot\nabla$ and the polarization velocity (\ref{eq:u_pol}) now appears explicitly in the term $d{\bf w}_{\bot}/dt$.

\section{\label{sec:dfecl}Local and Global Drift-fluid Energy Conservation Laws}

In this Section, we present the local and global forms of the energy conservation law, as they arise from an application of the Noether method \cite{Noether_1}-\cite{Noether_3}. For this purpose, we point out that, as a result of the variational principle $\int \delta\L\,d^{3}x\,dt = 0$, the only remaining terms in the Eulerian variation of the Lagrangian density (\ref{eq:Eul_Lag}) yield the Noether equation
\begin{equation}
\delta\L \;=\; \pd{\Lambda}{t} \;+\; \nabla\bdot\vb{\Gamma},
\label{eq:Noether}
\end{equation}
where the Noether fields $\Lambda$ and $\vb{\Gamma}$ are defined in Eqs.~(\ref{eq:Lambda})-(\ref{eq:Gamma}). The energy and momentum conservation laws are derived from the Noether equation (\ref{eq:Noether}) by considering infinitesimal time and space translations \cite{PCR_1}, respectively. In the present work, we focus our attention on the local and global energy conservation laws associated with our drift-Alfv\'{e}n model.

\subsection{Local energy conservation law}

We derive the local form of the energy conservation law from the Noether equation (\ref{eq:Noether}) by considering infinitesimal time translations $t \rightarrow t + \delta t$, from which we obtain the following expressions for the virtual fluid displacement $\x$ and the Eulerian variations $\delta\phi$ and $\delta\L$:
\begin{equation}
\x \;=\; -\,{\bf u}\,\delta t \;\;\;{\rm and}\;\;\; (\delta\phi,\; \delta A_{\|},\; \delta\L) \;=\; -\,\delta t\;\left( \pd{\phi}{t},\; \pd{A_{\|}}{t},\;
\pd{\L}{t}\right).
\label{eq:time}
\end{equation}
Inserting these expressions into Eqs.~(\ref{eq:Lambda})-(\ref{eq:Gamma}), the Noether fields become
\begin{eqnarray}
\Lambda & \equiv & -\;\delta t\;\left( {\bf u}\bdot\pd{\L}{{\bf u}} \right), 
\label{eq:Lambda_time} \\
\vb{\Gamma} & \equiv & -\;\delta t \left[\; {\bf u} \left( {\bf u}\bdot\pd{\L}{{\bf u}} \;-\; \eta^{a}\;
\pd{\L}{\eta^{a}} \right) \;+\; {\sf P}\bdot{\bf u} \;+\; \pd{\psi^{i}}{t}\;\pd{\L}{(\nabla\psi^{i})} \;\right], 
\label{eq:Gamma_time}
\end{eqnarray}
where summation over fluid species is, henceforth, implied wherever appropriate. By combining 
Eqs.~(\ref{eq:time})-(\ref{eq:Gamma_time}), we obtain the {\it primitive} form of the local energy conservation law:
\begin{equation} 
\pd{\varepsilon^{\prime}}{t} \;+\; \nabla\bdot{\bf S}^{\prime} \;=\; 0,
\label{eq:Econs_prim}
\end{equation}
where the primitive energy density is
\begin{eqnarray}
\varepsilon^{\prime} & \equiv & {\bf u}\bdot\pd{\L}{{\bf u}} \;-\; \L \nonumber \\ 
 & = & \frac{1}{2}\,nm\,U_{\|}^{2} \;+\; \left( p_{\bot} \;+\; \frac{p_{\|}}{2} \right) \;+\; en\,\Phi^{*} \;+\; \frac{1}{8\pi} \left( |\nabla_{\bot}A_{\|}|^{2} \;-\; |\nabla_{\bot}\phi|^{2} \right), \label{eq:E_prim}
\end{eqnarray}
and the primitive energy-density flux is
\begin{eqnarray}
{\bf S}^{\prime} & \equiv & {\bf u} \left( {\bf u}\bdot\pd{\L}{{\bf u}} \;-\; \eta^{a}\; \pd{\L}{\eta^{a}} \right) 
\;+\; {\sf P}\bdot{\bf u} \;+\; \pd{\psi^{i}}{t}\;\pd{\L}{(\nabla\psi^{i})} \nonumber \\
& = & \left[\; n \left(\frac{m}{2}\,U_{\|}^{2} \;+\; e\,\Phi^{*} \right) \;+\; \left( p_{\bot} \;+\; \frac{p_{\|}}{2} \right) \;\right] {\bf u} \;+\;{\sf P}\bdot{\bf u} \;+\; \pd{\psi^{i}}{t}\;\pd{\L}{(\nabla\psi^{i})}.
\label{eq:S_prim}
\end{eqnarray}
Note that the energy conservation law (\ref{eq:Econs_prim}) has the following gauge-invariance property: the energy conservation law is unaffected by the transformation 
\begin{equation}
\varepsilon \;=\; \varepsilon^{\prime} \;+\; \nabla\bdot{\bf D} \;\;\;{\rm and}\;\;\; {\bf S} \;=\; {\bf S}^{\prime} 
\;-\; \pd{{\bf D}}{t}, 
\label{eq:energy_gauge}
\end{equation}
where ${\bf D}$ is an arbitrary vector field.

In order to arrive at the final form of the local energy conservation law, we need to rearrange some terms in Eq.~(\ref{eq:E_prim}). First, by substituting the drift-fluid Poisson equation (\ref{eq:polarization}), we find
\begin{equation}
en\;\Phi^{*} \;-\; \frac{|\nabla\phi|^{2}}{8\pi} \;=\; \frac{|\nabla\phi|^{2}}{8\pi} \;+\; \frac{mn}{2}\;|{\bf u}_{E}|^{2} \;-\; mn\,u_{\|}\,V_{\|} \;-\; \nabla\bdot{\bf D},
\label{eq:switch_phi}
\end{equation}
where the gauge vector field ${\bf D}$ is defined as
\begin{equation}
{\bf D} \;\equiv\; \phi\;\pd{\L}{(\nabla\phi)} \;=\; -\;\phi\;\left( \frac{{\bf E}_{\bot}}{4\pi} \;+\; {\bf P}_{\bot} \right).
\label{eq:energy_gauge_D}
\end{equation}
Next, we express the primitive energy density (\ref{eq:E_prim}) as $\varepsilon^{\prime} \equiv \varepsilon \;-\; \nabla\bdot{\bf D}$, where the final form of the energy density is defined as
\begin{equation}
\varepsilon \;\equiv\; \frac{1}{2}\,mn\; \left| u_{\|}\,(\bhat_{0} + {\bf B}_{\bot}/B_{0}) + {\bf u}_{E}\right|^{2} \;+\; 
\left( p_{\bot} \;+\; \frac{p_{\|}}{2} \right) \;+\; \frac{1}{8\pi} \left( |{\bf E}_{\bot}|^{2} \;+\; 
|{\bf B}_{\bot}|^{2} \right).
\label{eq:energy_density}
\end{equation}
The final form of the local energy conservation law is, therefore, expressed as
\begin{equation}
\pd{\varepsilon}{t} \;+\; \nabla\bdot{\bf S} \;=\; 0,
\label{eq:ECL}
\end{equation}
where the final form of the energy density flux, defined as
\begin{equation}
{\bf S} \;\equiv\; {\bf S}^{\prime} \;-\; \pd{{\bf D}}{t},
\label{eq:energy_flux}
\end{equation}
is expressed, after some partial cancelations, as
\begin{equation}
{\bf S} \;=\; \left[\; n \left(\frac{m}{2}\;U_{\|}^{2} \;+\; e\,\Phi^{*} \right) \;+\; \left( p_{\bot} \;+\; 
\frac{p_{\|}}{2} \right) \;\right] {\bf u} \;+\; {\sf P}\bdot{\bf u} \;+\; {\bf S}_{\varphi},
\label{eq:S_final}
\end{equation}
where the field energy density flux ${\bf S}_{\varphi}$ is defined as
\begin{equation} 
{\bf S}_{\varphi} \;\equiv\; -\; \phi\;\pd{}{t}\left( \pd{\L}{(\nabla\phi)}\right) \;+\; \pd{A_{\|}}{t}\;\pd{\L}{(\nabla A_{\|})}.
\label{eq:electric_flux}
\end{equation}

In Appendix B, after gyroviscous diamagnetic cancellations are inserted into the drift-fluid equation for 
$(u_{\|},p_{\|},p_{\bot})$, the local energy conservation law (\ref{eq:ECL}) is converted into a new energy equation 
\begin{equation}
\pd{\varepsilon}{t} \;+\; \nabla\bdot{\bf S}^{*} \;=\; 0, 
\label{eq:energy_dia}
\end{equation}
in which diamagnetic-cancellation terms result in a heat-modified energy density flux ${\bf S}^{*}$ [see 
Eq.~(\ref{eq:S_star})]. This new form ensures that the total energy ${\cal E} = \int \varepsilon\,d^{3}x$ still satisfies 
the global energy conservation law $d{\cal E}/dt = 0$.

\subsection{Global energy conservation law}

The global energy conservation law 
\begin{equation}
0 \;=\; \frac{d{\cal E}}{dt} \;\equiv\; \left. \left. \frac{d}{dt} \right( {\cal E}_{u} \;+\; {\cal E}_{p} \;+\; 
{\cal E}_{\phi} \;+\; {\cal E}_{A} \right),
\label{eq:energy_global} 
\end{equation}
can be derived from the local energy conservation law (\ref{eq:ECL}) by integrating it over space (and neglecting surface terms), where the component energies
\begin{eqnarray}
{\cal E}_{u} & = & \int d^{3}x\; \left( \frac{mn}{2}\,U_{\|}^{2} \right), \label{eq:E_u} \\
{\cal E}_{p} & = & \int d^{3}x\; \left( p_{\bot} \;+\; \frac{p_{\|}}{2} \right), \label{eq:E_p} \\ 
{\cal E}_{\phi} & = & \int d^{3}x\; \left( \frac{|\nabla\phi|^{2}}{8\pi} \;+\; \frac{mn}{2}\,|{\bf u}_{E}|^{2} \;-\; 
mnu_{\|}\,V_{\|} \right), \label{eq:E_phi} \\
{\cal E}_{A} & = & \int d^{3}x\; \left( \frac{|\nabla_{\bot}A_{\|}|^{2}}{8\pi} \right),
\label{eq:E_A}
\end{eqnarray}
represent the parallel kinetic energy, the internal energy, and the electromagnetic field energy (defined as the sum of the $\phi$-field energy ${\cal E}_{\phi}$ and the $A_{\|}$-field energy ${\cal E}_{A}$), respectively. We now present the time evolution of each energy component (in the absence of diamagnetic cancellations and heat fluxes) in order to identify the energy-exchange processes that allow the transfer of energy between the three types of drift-fluid (parallel kinetic, internal, electromagnetic field) energies. 

First, we begin with the time derivative of the parallel kinetic energy density
\begin{eqnarray}
\pd{}{t}\left( \frac{mn}{2}\,U_{\|}^{2} \right) & = & -\;\nabla\bdot\left[\; \left( \frac{mn}{2}\,U_{\|}^{2} \right)
\;{\bf u}\;\right] \;-\; {\bf u}\bdot\left( \nabla\bdot{\sf P} \;+\; en \nabla\Phi^{*} \right) \nonumber \\
 &  &\mbox{}-\; n\,u_{\|} \left( mu_{\|}\;b\,\pd{b}{t} \;+\; \frac{e}{c}\;\pd{A_{\|}^{*}}{t} \right),
\label{eq:E_u_first}
\end{eqnarray}
derived from Eq.~(\ref{eq:u_parallel}) after rearranging some terms and using the continuity equation 
(\ref{eq:continuity}). Hence, the time evolution of the parallel-kinetic energy is
\begin{equation}
\frac{d{\cal E}_{u}}{dt} \;=\; -\;\int d^{3}x\; \left[\; {\bf u}\bdot\left( \nabla\bdot{\sf P} \;+\; en \nabla\Phi^{*}\right) \;+\; n\,u_{\|} \left( mu_{\|}\;b\,\pd{b}{t} \;+\; \frac{e}{c}\;\pd{A_{\|}^{*}}{t} \right) \;\right].
\label{eq:E_u_time}
\end{equation}

Next, we evaluate the time derivative of the internal energy density $\P \equiv p_{\bot} + p_{\|}/2$ using the parallel and perpepdicular CGL pressure equations (\ref{eq:pperp_CGL})-(\ref{eq:ppar_CGL}):
\[ \frac{d\P}{dt} \;+\; \P\;\nabla\bdot{\bf u} \;+\; {\sf P}:\nabla{\bf u} \;=\; 0, \]
so that, by rearranging terms, we obtain
\begin{equation}
\pd{\P}{t} \;=\; -\;\nabla\bdot\left( \P\;{\bf u} \;+\; {\sf P}\bdot{\bf u} \right) \;+\; {\bf u}\bdot
\left(\nabla\bdot{\sf P}\right),
\label{eq:E_p_first}
\end{equation}
and the time evolution of the internal energy is
\begin{equation}
\frac{d{\cal E}_{p}}{dt} \;=\; \int d^{3}x\; {\bf u}\bdot\left(\nabla\bdot{\sf P} \right).
\label{eq:E_p_time}
\end{equation}

Thirdly, we evaluate the time derivative of the $A_{\|}$-field energy density
\begin{eqnarray}
\pd{}{t} \left(\frac{|\nabla_{\bot}A_{\|}|^{2}}{8\pi} \right) & = & -\;\frac{1}{4\pi}\,\pd{A_{\|}}{t}\;
\nabla_{\bot}^{2}A_{\|} \;+\; \nabla\bdot\left( \pd{A_{\|}}{t}\; \frac{\nabla_{\bot}A_{\|}}{4\pi} \right) \nonumber \\
 & = & -\;\nabla\bdot\left( \pd{A_{\|}}{t}\;\pd{\L}{\nabla A_{\|}} \right) \;+\; \frac{en}{c}\,u_{\|}\;\pd{A_{\|}}{t} 
\nonumber \\
 &  &-\; \frac{mnu_{\|}c}{B_{0}^{2}} \left( \nabla_{\bot}\phi \;-\; \frac{u_{\|}}{c}\;\nabla_{\bot}A_{\|} \right)\bdot\left(\nabla_{\bot}\pd{A_{\|}}{t}\right),
\label{eq:E_A_first}
\end{eqnarray}
where the second equality is obtained by substituting the drift-fluid Amp\`{e}re equation (\ref{eq:A_Maxwell}), so that the time evolution of the $A_{\|}$-field energy is
\begin{equation}
\frac{d{\cal E}_{A}}{dt} \;=\; \int d^{3}x \left[\; nu_{\|} \left( mu_{\|}\;b\,\pd{b}{t} \;+\; 
\frac{e}{c}\;\pd{A_{\|}}{t} \right) \;-\; \frac{mnu_{\|}c}{B_{0}^{2}}\;\nabla_{\bot}\phi\bdot
\left(\nabla_{\bot}\pd{A_{\|}}{t}\right) \;\right].
\label{eq:E_A_time}
\end{equation}

Lastly, we evaluate time derivative of the $\phi$-field energy density
\begin{eqnarray}
\pd{}{t}\left( \frac{|\nabla_{\bot}\phi|^{2}}{8\pi} \;+\; \frac{mn}{2}\,|{\bf u}_{E}|^{2} \;-\; mnu_{\|}\,V_{\|} \right) 
& = & \pd{}{t} \left( en\,\Phi^{*} \;-\; \frac{|\nabla_{\bot}\phi|^{2}}{8\pi} \;+\; \nabla\bdot{\bf D} \right)  
\nonumber \\
 & = & \nabla\bdot\left( \pd{{\bf D}}{t} \;-\; {\bf u}\;en\,\Phi^{*} \right) \;+\; {\bf u}\bdot\left( en\,\nabla\Phi^{*}\right) \nonumber \\
 &  &+\; \left( en\;\pd{\Phi^{*}}{t} \;-\; \frac{\nabla_{\bot}\phi}{4\pi}\bdot\nabla_{\bot}\pd{\phi}{t} \right),
\label{eq:E_phi_first}
\end{eqnarray}
where the first equality is obtained from the definition (\ref{eq:switch_phi}) while the second equality arises from the use of the continuity equation (\ref{eq:continuity}). By using the definition 
\begin{eqnarray*}
en\;\pd{\Phi^{*}}{t} \;-\; \frac{\nabla_{\bot}\phi}{4\pi}\bdot\nabla_{\bot}\pd{\phi}{t} & = & \pd{\phi}{t} \left[\; en 
\;+\; \nabla\bdot\left( \frac{\nabla\phi}{4\pi} \;+\; \frac{mnc^{2}}{B_{0}^{2}}\;\nabla_{\bot}\phi \right) \;\right] \\
 &  &\mbox{}-\; \nabla\bdot\left[\; \pd{\phi}{t} \left( \frac{\nabla_{\bot}\phi}{4\pi} \;+\; \frac{mnc^{2}}{B_{0}^{2}}\;
\nabla_{\bot}\phi \right) \;\right] \\
 & = & -\;\nabla\bdot\left( \pd{\phi}{t}\;\pd{\L}{\nabla\phi} \right) \;-\; mnu_{\|}\;\left(\frac{c}{B_{0}^{2}}\,
\nabla_{\bot}\pd{\phi}{t}\right)\bdot\nabla_{\bot}A_{\|},
\end{eqnarray*}
so that Eq.~(\ref{eq:E_phi_first}) becomes
\begin{eqnarray}
 &  & \pd{}{t}\left( \frac{|\nabla_{\bot}\phi|^{2}}{8\pi} \;+\; \frac{mn}{2}\,|{\bf u}_{E}|^{2} \;-\; mnu_{\|}\,V_{\|}  \right) \;=\; \nabla\bdot\left[\; \phi\;\pd{}{t}\left(\pd{\L}{\nabla\phi}\right) \;-\; en\,\Phi^{*}\;{\bf u} \;\right]  \nonumber \\ 
 &  &\;\;\;+\; {\bf u}\bdot\left( en\,\nabla\Phi^{*}\right) \;-\; mnu_{\|}\;\left( \frac{c}{B_{0}^{2}}\,\nabla_{\bot}\pd{\phi}{t}\right) \bdot\nabla_{\bot}A_{\|},
\label{eq:E_phi_second}
\end{eqnarray}
and the time evolution of the $\phi$-field energy is
\begin{equation}
\frac{d{\cal E}_{\phi}}{dt} \;=\; \int d^{3}x \left[\; {\bf u}\bdot\left( en\,\nabla\Phi^{*}\right) \;-\; mnu_{\|}\;
\left( \frac{c}{B_{0}^{2}}\,\nabla_{\bot}\pd{\phi}{t}\right)\bdot\nabla_{\bot}A_{\|} \;\right].
\label{eq:E_phi_time}
\end{equation}

By combining Eqs.~(\ref{eq:E_A_time}) and (\ref{eq:E_phi_time}), and using the expression
\[ \pd{V_{\|}}{t} \;=\; \frac{c}{B_{0}^{2}} \left( \nabla_{\bot}\pd{\phi}{t}\bdot\nabla_{\bot}A_{\|} \;+\; \nabla_{\bot}\phi\bdot\nabla_{\bot}
\pd{A_{\|}}{t} \right) \]
with the definition (\ref{eq:phi_Ab}) for $A_{\|}^{*}$, we obtain the time evolution of the electromagnetic energy
${\cal E}_{\psi} = {\cal E}_{\phi} + {\cal E}_{A}$:
\begin{equation}
\frac{d{\cal E}_{\psi}}{dt} \;=\; \int d^{3}x \left[\; en{\bf u}\bdot\nabla\Phi^{*} \;+\; nu_{\|} \left( mu_{\|}\;b\,
\pd{b}{t} \;+\; \frac{e}{c}\,\pd{A_{\|}^{*}}{t} \right) \;\right].
\label{eq:EM_energy}
\end{equation}
We now see that by combining Eqs.~(\ref{eq:E_u_time}), (\ref{eq:E_p_time}), and (\ref{eq:EM_energy}), the global energy conservation law (\ref{eq:energy_global}) is satisfied exactly. Hence, the energy-transfer processes (\ref{eq:E_u_time}), (\ref{eq:E_p_time}), and (\ref{eq:EM_energy}) between the parallel kinetic energy ${\cal E}_{u}$, internal energy 
${\cal E}_{p}$, and electromagnetic field energy ${\cal E}_{\psi}$ involve the following pathways:
\begin{equation}
(p_{\bot},p_{\|}) \;\;\;\leftrightarrow\;\;\; (n,u_{\|}) \;\;\;\leftrightarrow\;\;\; (\phi,A_{\|}),
\label{eq:energy_transfer}
\end{equation}
where Eqs.~(\ref{eq:E_u_time}) and (\ref{eq:E_p_time}), for example, show that internal energy ${\cal E}_{p}$ is transferred to and from parallel kinetic energy ${\cal E}_{u}$.

\section{\label{sec:disc}Summary}

The nonlinear drift-Alfv\'{e}n equations (\ref{eq:continuity}), (\ref{eq:u_parallel}), and 
(\ref{eq:pperp_CGL})-(\ref{eq:ppar_CGL}) for the anisotropic drift-fluid moments $(n,u_{\|},p_{\bot},p_{\|})$, respectively, and the drift-fluid Poisson-Amp\`{e}re equations (\ref{eq:polarization}) and (\ref{eq:A_Maxwell}) were derived in this paper from a Lagrangian variational principle based on the drift-fluid Lagrangian density 
(\ref{eq:Lag_total}). The local and global forms of the exact energy conservation law for these drift-Alfv\'{e}n equations, given by Eqs.~(\ref{eq:ECL}) and (\ref{eq:energy_global}), respectively, were derived by the Noether 
method as a natural consequence of the variational formulation presented here.

The inclusion of the linear and nonlinear $E\times B$ velocities (\ref{eq:ExB_lin})-(\ref{eq:ExB_nonlin}) as well as the parallel flow along the total magnetic field lines into the drift-fluid Lagrangian density (\ref{eq:Lag_total}) enabled a derivation of generalized (electromagnetic) drift-fluid polarization and magnetization effects. Under the drift-fluid quasineutrality condition, the drift-fluid Poisson-Amp\`{e}re equations (\ref{eq:polarization}) and (\ref{eq:A_Maxwell}) were transformed into the drift-fluid charge conservation law (\ref{eq:df_charge}), with the net current density 
${\bf J}$ defined by Eq.~(\ref{eq:Jnet_def}), which represents a generalized (electromagnetic) form of the so-called vorticity equation.

Although higher-order drift-fluid moments have been omitted from the present variational formulation of our four-moment nonlinear anisotropic drift-Alfv\'{e}n model, Appendix A shows how higher-order heat fluxes might be introduced into a variational formulation, while Appendix B shows how diamagnetic cancellations in the drift-fluid equations for 
$(u_{\|},p_{\bot},p_{\|})$ can be incorporated without affected the global energy conservation law. Future work will investigate the extension of variational principles by including higher-order heat fluxes as dynamical variational fields as well as the extension of Generalized Lagrangian Mean (GLM) methods \cite{GLM} to plasma fluid models.

\acknowledgments

The Author wishes to acknowledge Drs.~Bruce Scott, T.~S.~Hahm, and R.~E.~Denton, for their useful comments and continued interest in this work.

The present work was supported by the National Science Foundation under grant No.~DMS-0317339.

\appendix

\section{Notes on Macroscopic Plasma Lagrangians}

In this Appendix, we present an adapted version of Y.-K.~M.~Peng's Ph.~D.~dissertation \cite{Peng_PhD} (with selected parts also published in Ref.~\cite{Peng_Crawford}). The macroscopic Lagrangian presented by Peng is based on constrained variations of the fluid density, velocity, and pressure tensor, and includes the effects of finite heat fluxes self-consistently.

\subsection{Lagrangian Density in Eulerian Coordinates}

Following Peng's work, we begin by combining the pressure constraint equations (\ref{eq:pperp_CGL})-(\ref{eq:ppar_CGL}) to form the scalar pressure equation for $\P \equiv p_{\bot} + p_{\|}/2$:
\begin{equation}
\frac{d\P}{dt} \;+\; \P\;\nabla\bdot{\bf u} \;+\; {\sf P}:\nabla{\bf u} \;=\; -\;\nabla\bdot{\bf q},
\label{eq:CGL_heat}
\end{equation}
where the heat flux ${\bf q}$ now appears on the right side of Eq.~(\ref{eq:CGL_heat}). Next, the heat flux vector 
${\bf q}$ is represented as
\begin{equation} 
{\bf q} \;\equiv\; \frac{d{\bf Q}}{dt} \;=\; \pd{{\bf Q}}{t} \;+\; {\bf u}\bdot\nabla{\bf Q},
\label{eq:qQ_def}
\end{equation}
where ${\bf Q}$ denotes the total density of heat energy transported across the boundary of the macroscopic (coarse-graining) cell as it moves along its Lagrangian trajectory \cite{Peng_PhD,Peng_Crawford}. Hence, the right side of Eq.~(\ref{eq:CGL_heat}) can be written as
\[ -\;\nabla\bdot{\bf q} \;\equiv\; -\;\nabla\bdot\frac{d{\bf Q}}{dt} \;=\; -\;\frac{d\nabla\bdot{\bf Q}}{dt} \;-\; 
\nabla{\bf u}:\nabla{\bf Q}, \]
so that Eq.~(\ref{eq:CGL_heat}) may be rewritten as
\begin{equation}
\frac{d}{dt} \left( \P \;+\; \nabla\bdot{\bf Q} \right) \;=\; -\;\P\;\nabla\bdot{\bf u} \;-\; \left( {\sf P} 
\;+\; \nabla{\bf Q} \right):\nabla{\bf u}.
\label{eq:CGL_Q}
\end{equation}
By using the Lagrangian-variation limitting procedure (\ref{eq:Lagvar_def}), we, thus, obtain the Lagrangian variation
\begin{equation} 
\Delta \left( \P \;+\; \nabla\bdot{\bf Q} \right) \;=\; -\;\P\;\nabla\bdot\vb{\xi} \;-\; \left( {\sf P} \;+\; 
\nabla{\bf Q} \right):\nabla\vb{\xi},
\label{eq:pressure_lag}
\end{equation}
from which we obtain the Eulerian variation
\begin{equation}
\delta\left( \P \;+\; \nabla\bdot{\bf Q} \right) \;=\; -\;\nabla\bdot\left( \P\;\vb{\xi} \;+\; \vb{\xi}\bdot\nabla{\bf Q} \right) \;-\; {\sf P}:\nabla\vb{\xi}.
\label{eq:pressure_var}
\end{equation}
Lastly, the heat flux ${\bf q}$ is introduced in the Lagrangian density $\L$, defined by Eq.~(\ref{eq:Lag_total}), by adding the divergence term $-\,\nabla\bdot{\bf Q}$:
\begin{equation}
\L^{\prime} \;\equiv\; \L \;-\; \nabla\bdot{\bf Q} \;\equiv\; \L_{cold} \;-\; \left( \P \;+\; \nabla\bdot{\bf Q} \right),
\label{eq:L_Q}
\end{equation}
where $\L_{cold}$ denotes the cold-fluid part of the drift-fluid Lagrangian density (\ref{eq:Lag_total}).

\subsection{Eulerian Variational Principle}

We now write the expression for the Eulerian variation of the heat-modified Lagrangian density (\ref{eq:L_Q}):
\begin{equation}
\delta\L^{\prime} \;=\; \delta\L_{cold} \;+\; \nabla\bdot\left( \P\;\vb{\xi} \;+\; 
\vb{\xi}\bdot\nabla{\bf Q} \right) \;+\; {\sf P}:\nabla\vb{\xi} \;\equiv\; \delta\L \;+\; \nabla\bdot\left( 
\vb{\xi}\bdot\nabla{\bf Q} \right),
\label{eq:delta_Lq}
\end{equation}
where the Eulerian variation $\delta\L$ is given by the Noether equation (\ref{eq:Noether}). We note that the quantity 
${\bf Q}$ does not play a dynamic role since it appears inside a spatial divergence. We shall see, however, that it plays an important role in terms of the energy conservation law derived from the heat-modified Noether equation
\begin{equation}
\delta\L^{\prime} \;=\; \pd{\Lambda}{t} \;+\; \nabla\bdot\left( \vb{\Gamma} \;+\; \x\bdot\nabla{\bf Q} \right).
\label{eq:Noether_Q}
\end{equation}

\subsection{Energy Conservation Law by Noether Method}

The invariance of the Lagrangian density (\ref{eq:L_Q}) under time translations $t \rightarrow t + \delta t$ yields an energy equation derived from the Noether equation (\ref{eq:Noether_Q}). By applying the Noether method (as outlined in Sec.~\ref{sec:dfecl}), we obtain the heat-modified energy equation:
\begin{equation}
0 \;=\; \pd{}{t} \left( \varepsilon \;+\; \nabla\bdot{\bf Q} \right) \;+\; \nabla\bdot\left( {\bf S} \;+\; {\bf u}\bdot\nabla{\bf Q} \right) \;\equiv\; \pd{\varepsilon}{t} \;+\; \nabla\bdot\left( {\bf S} \;+\; 
\frac{d{\bf Q}}{dt} \right),
\label{eq:ECL_Q}
\end{equation}
where the energy density $\varepsilon$ and energy-density flux ${\bf S}$ are defined in Eqs.~(\ref{eq:energy_density}) 
and (\ref{eq:S_final}), respectively. Lastly, using the definition (\ref{eq:qQ_def}) for the heat flux ${\bf q}$, we obtain the energy equation
\begin{equation}
\pd{\varepsilon}{t} \;+\; \nabla\bdot{\bf S} \;=\; -\;\nabla\bdot{\bf q}.
\label{eq:energy_heat}
\end{equation}

Although the heat flux ${\bf q}$ appears self-consistently within Peng's variational principle, we note that the evolution equation for ${\bf q}$ is outside of its scope. It is a topic of future research to determine whether the evolution equation for the heat flux can be derived within a variational formulation.

\section{\label{sec:dia_can}Diamagnetic cancelations and Energy conservation}

The drift-fluid equations (\ref{eq:u_parallel}) and (\ref{eq:pperp_CGL})-(\ref{eq:ppar_CGL}) contain diamagnetic advective derivatives ${\bf u}_{D}\bdot\nabla\eta^{a}$ of the drift-fluid moments $\eta^{a} = (u_{\|},p_{\bot},p_{\|})$ that must be cancelled by the addition of so-called higher-order gyroviscous cancellations. Since these higher-order moments cannot be derived from the drift-fluid Lagrangian, the diamagnetic cancelations must be inserted in drift-fluid equations in the post-variational phase. However, there exists a constraint in the addition of higher-order moment terms, namely, that the global energy conservation law should not be altered. 

\subsection{Parallel drift-fluid dynamics}

The diamagnetic cancelation needed for the parallel momentum equation (\ref{eq:u_parallel}) involves the addition of the term $-\,\nabla\bdot\vb{\Pi}_{\|}^{*}$ on the right side of Eq.~(\ref{eq:u_parallel}), associated with the non-diagonal part of the pressure tensor. From Vlasov theory, the diamagnetic-cancelation term is found to be 
\cite{Braginskii,HM,Chang_Callen}:
\begin{equation}
-\;\nabla \bdot \vb{\Pi}_{\|}^{*} \;=\; -\;\nabla \bdot \left( p_{\perp}\; \frac{mc\bhat_{0}}{eB_{\|}^{*}}\btimes 
\nabla u_{\|} \right) \;=\; mn{\bf u}_{D}\bdot\nabla u_{\|} \;+\; p_{\bot}\;\K(mu_{\|}),
\label{eq:diacan_upar}
\end{equation}
where the term $B_{\|}^{*}$ replaces the denominator $B_{0}$ to ensure an exact diamagnetic cancellation and the 
{\it magnetic} differential operator $\K(\cdots)$ is defined by the identity
\begin{equation} 
\nabla\bdot\left( g\;\frac{c\bhat_{0}}{eB_{\|}^{*}}\btimes\nabla f \right) \;=\; \frac{c\bhat_{0}}{eB_{0}}\bdot\nabla f\btimes\nabla g \;-\; g\,\K(f),
\label{eq:K_id}
\end{equation}
valid for arbitrary functions $f$ and $g$.

Note that the diamagnetic-cancellation term (\ref{eq:diacan_upar}) is energy conserving since
\[ -\;u_{\|}\;\nabla \bdot \vb{\Pi}_{\|}^{*} \;\equiv\; -\;\nabla \bdot\left( u_{\|}\;\vb{\Pi}_{\|}^{*} \right) \]
and, thus, the time evolution of the parallel kinetic energy density (\ref{eq:E_u_first}) is now expressed as
\begin{eqnarray}
\pd{}{t} \left(\frac{mn}{2}\;U_{\|}^{2} \right) & = & -\;\nabla\bdot \left( \frac{mn}{2}\;U_{\|}^{2}\,{\bf u} \;+\; 
u_{\|}\;\vb{\Pi}_{\|}^{*} \right) \;-\; {\bf u}\bdot\left( \nabla\bdot{\sf P} \;+\; en\nabla\Phi^{*} \right) \nonumber \\
 &  &\mbox{}-\; n u_{\|}\; \left( mu_{\|}\;b\,\pd{b}{t} \;+\; \frac{e}{c}\;\pd{A_{\|}^{*}}{t} \right),
\label{eq:par_kinetic}
\end{eqnarray}
and the global time evolution equation (\ref{eq:E_u_time}) for ${\cal E}_{u}$ is still valid.

\subsection{Internal energy}

To consider the diamagnetic cancellations in the anisotropic pressure equations (\ref{eq:pperp_CGL})-(\ref{eq:ppar_CGL}), we write the modified CGL pressure equations \cite{weiland}:
\begin{eqnarray}
\frac{dp_{\bot}}{dt} \;+\; p_{\bot}\;\nabla\bdot{\bf u} \;+\; p_{\perp}\;({\bf I} - \bhat_{0}\bhat_{0}):\nabla{\bf u} 
& = & -\;\nabla\bdot{\bf q}_{\perp}^{(\bot)} \;-\; 2\,{\bf q}_{\perp}^{(\|)}\bdot (\bhat_{0}\bdot \nabla\bhat_{0}), 
\label{eq:p_perpq} \\
\frac{dp_{\|}}{dt} \;+\; p_{\|}\;\nabla\bdot{\bf u} \;+\; 2\,p_{\parallel}\;\bhat_{0}\bhat_{0}:\nabla{\bf u} & = & -\;2\,\nabla\bdot{\bf q}_{\perp}^{(\|)} \;+\; 4\,{\bf q}_{\perp}^{(\|)}\bdot (\bhat_{0}\bdot \nabla\bhat_{0}), 
\label{eq:p_parq}
\end{eqnarray}
which have been modified by the addition of the parallel and perpendicular gyroviscous heat fluxes 
${\bf q}_{\perp}^{(\|)}$ and ${\bf q}_{\perp}^{(\bot)}$, respectively, derived from the Vlasov equation directly and found to be
\begin{equation}
{\bf q}_{\perp}^{(\|)} \;=\; \frac{1}{2}\;\frac{cp_{\perp}\bhat_{0}}{eB_{\|}^{*}} \btimes \nabla T_{\parallel} \;+\; 
p_{\|}\;{\bf u}_{C} \;\;\;{\rm and}\;\;\; {\bf q}_{\perp}^{(\bot)} \;=\; 2\;\frac{cp_{\perp}\bhat_{0}}{eB_{\|}^{*}}
\btimes \nabla T_{\perp}.
\label{eq:q_weiland}
\end{equation}
Inserting these diamagnetic heat fluxes into Eqs.~(\ref{eq:p_perpq})-(\ref{eq:p_parq}), the time evolution of the internal energy can, therefore, be expressed as
\begin{equation}
\pd{\P}{t} \;=\; -\;\nabla\bdot\left( \P\;{\bf u} \;+\; {\sf P}\bdot{\bf u} \;+\; {\bf q} \right) \;+\; 
{\bf u}\bdot\left(\nabla\bdot{\sf P}\right),
\label{eq:internal}
\end{equation}
where the net heat flux is ${\bf q} \equiv {\bf q}_{\perp}^{(\|)} + {\bf q}_{\perp}^{(\bot)}$. Hence, we see that the insertion of diamagnetic cancellations into our finite-$\beta$ electromagnetic drift-fluid model does not jeopardize the local and global energy conservation laws, where the energy density flux (\ref{eq:S_final}) is now replaced by the flux
\begin{equation} 
{\bf S}^{*} \;=\; {\bf S} \;+\; u_{\|}\;\vb{\Pi}_{\|}^{*} \;+\; {\bf q}_{\perp}^{(\|)} \;+\; {\bf q}_{\perp}^{(\bot)}.
\label{eq:S_star}
\end{equation}

\end{document}